\documentclass{emulateapj}
\shorttitle{KELVIN-HELMHOLTZ INSTABILITY WITH COSMIC-RAYS}
\shortauthors{SUZUKI,TAKAHASHI\&KUDOH}
\begin{document}
\title{LINEAR GROWTH OF THE KELVIN-HELMHOLTZ INSTABILITY WITH AN ADIABATIC COSMIC-RAY GAS}
\author{AKIHIRO SUZUKI\altaffilmark{1,4}, HIROYUKI R. TAKAHASHI\altaffilmark{1}, and TAKAHIRO KUDOH\altaffilmark{2,3}}
\altaffiltext{1}{Center for Computational Astrophysics, National Astronomical Observatory of Japan, Mitaka, Tokyo, 181-8588, Japan.}
\altaffiltext{2}{Division of Theoretical Astrophysics, National Astronomical Observatory of Japan, Mitaka, Tokyo, 181-8588, Japan.}
\altaffiltext{3}{Department of Astronomical Science, School of Physical Sciences, the
Graduate University for Advanced Studies (SOKENDAI), Osawa 2-21-1,
Mitaka, Tokyo 181-8588, Japan.}
\altaffiltext{4}{Current address: Department of Astronomy, Kyoto University, Kitashirakawa-Oiwake-cho, Sakyo-ku, Kyoto 606-8502, Japan.}

\begin{abstract}
We investigate effects of cosmic-rays on the linear growth of the Kelvin-Helmholtz instability. 
Cosmic-rays are treated as an adiabatic gas and allowed to diffuse along magnetic field lines. 
We calculated the dispersion relation of the instability for various sets of two free parameters, the ratio of the cosmic-ray pressure to the thermal gas pressure and the diffusion coefficient. 
Including cosmic-ray effects, a shear layer is more destabilized and the growth rates can be enhanced in comparison with the ideal magnetohydrodynamical case. 
Whether the growth rate is effectively enhanced or not depends on the diffusion coefficient of cosmic-rays. 
We obtain the criterion for effective enhancement by comparing the growing time scale of the instability with the diffusion time scale of cosmic-rays. 
These results can be applied to various astrophysical phenomena where a velocity shear is present, such as outflows from star-forming galaxies, AGN jet, channel flows resulting from the nonlinear development of the magnetorotational instability, and galactic disks.
\end{abstract}

\section{INTRODUCTION\label{intro}}
Cosmic-rays, thermal gases, and magnetic fields are ubiquitous in the universe and indispensable components of astrophysical objects in various scales. 
How they interact with each other and how they exchange their energies are thought to key ingredients to understand various astrophysical phenomena. 
However, current understanding of the interaction between these components is not sufficient. 
There are numerous works considering how cosmic-rays affect thermal gas and magnetic fields in different ways. 
Some treat interactions between cosmic-rays and magnetohydrodynamics (MHD) waves in kinetic approach and others use the cosmic-ray transport equation in the diffusion limit \citep{1971ApJ...170..265S,1975MNRAS.172..557S}. 
Among them, one of the most convenient ways is to regard cosmic-rays as a gas composed of relativistic particles and allow the gas diffuse in the ISM with a specific diffusion coefficient \citep{1985A&A...151..151S}. 
Especially, in the presence of an ordered magnetic field, cosmic-rays diffuse along the magnetic field line since the gyration of cosmic-ray particles around the magnetic field line prevents efficient exchanges of energies between cosmic-ray particles \citep{1999ApJ...520..204G}. 
In this simplified manner, a lot of works have been done on the purpose of revealing how cosmic-ray pressure affects the dynamical evolution of the ISM. 

Cosmic-rays are considered to be one of major components of the interstellar medium(ISM). 
The energy density of cosmic-rays in the interstellar space is known to be comparable to those of thermal gas and magnetic fields. 
The considerably high energy density means that we should take into account effects of cosmic-rays on thermal gas and magnetic fields when we consider the dynamical evolution of the ISM. 
Also, in hot plasmas trapped in the gravitational potential of clusters of galaxies, i.e., the intra-cluster medium (ICM), cosmic-rays are considered to play important roles in transporting energy from AGN jets to surrounding materials.

For example, \cite{2003ApJ...589..338R} carried out the linear analysis of the Parker instability \citep{1966ApJ...145..811P,1992ApJ...401..137P} for various sets of the diffusion coefficient and the ratio of the cosmic-ray gas pressure to the thermal gas pressure. 
\cite{2003A&A...412..331H} and \cite{2004ApJ...607..828K} developed numerical codes solving MHD equations coupled with cosmic-ray pressure and investigated the linear and nonlinear development of the Parker instability in the presence of cosmic-rays. 
Later, simulations of the Galactic disk were performed by \cite{2004ApJ...605L..33H,2009A&A...498..335H} to reveal roles of cosmic-rays in driving the magnetohydrodynamical dynamo. 
Furthermore, effects of cosmic-rays on the linear growth of magnetorotational instability \citep{Velikhov1959,1960PNAS...46..253C,1991ApJ...376..214B} have been investigated by \cite{2012Ap&SS.337..247K}. 
Cosmic-rays possibly affect growth of the thermal instability \citep{1965ApJ...142..531F}. 
There are some studies on linear analysis of the instability \citep{1994ApJ...431..689B,2005A&A...430..567W,2009MNRAS.397.1521S}. 
In \cite{2010ApJ...720..652S}, linear analysis and numerical simulations on the thermal instability with anisotropic conduction and cosmic-ray gas transport are applied to the ICM.

In these systems, velocity shears would be naturally formed, suggesting development of Kelvin-Helmholtz(KH) instability. 
The KH instability is a fundamental process of hydrodynamical instability and thus is greatly paid attention in astrophysics \cite[e.g.,][]{1961hhs..book.....C}. 
For example, in a cluster of galaxies, member galaxies of the cluster move in the surrounding hot plasma. 
Therefore, the contact surface exists between the ICM and the ISM of each galaxy. 
At the contact surface, the KH instability can develop and the ISM may be stripped from the galaxy. 
The vortex acts as an amplifier of magnetic fields \citep{2007ApJ...663..816A,2013ApJ...768..175S}.
Shear flows can also be found in outflows in the ISM, the inter-galactic medium, and the ICM. 
The difference in the velocities of the jet and the ambient medium leads to the development of the KH instability. 
As a result, the jet material and the ambient medium are expected to be efficiently mixed up. 

Another example at which the KH instability plays important role is the accretion disks threaded by the magnetic fields. 
In such systems, gravitational energy of accreting gas is converted into the magnetic and kinetic energy through the MRI, resulting a formation of so-called channel flow. 
Although the channel flow structure is a fully exact solution of non-linear MHD equations and MRI is expected to continue to grow exponentially, the channel flow is actually disrupted by the secondary KH instability \citep{1994ApJ...432..213G}. 
Then the disk becomes fully turbulent system. 
Thus, the saturation levels of MRI mode and rate of angular momentum transport would be strongly affected by the growth of the KH instability. 
Since saturation levels of MRI determines the amount of the magnetic energy transported into the disk corona, the growth of the KH instability would impact on the coronal heating \citep{2012arXiv1205.6537U}. 
Also we have to note that the KH instability is important to consider the disk wind from the turbulent accretion disks. 
The accretion disks threaded by the magnetic field can drive disk wind \citep{1982MNRAS.199..883B}. 
The disk winds driven from the turbulent accretion disks suffer from a kind of the KH instability \citep{2013A&A...550A..61L} and resulting in the formation of turbulent outflows \citep{2013A&A...552A..71F}.  

Cosmic-rays are thought to have a potential to drive outflows from star-forming galaxies, i.e., so-called cosmic-ray driven wind \citep{1975ApJ...196..107I}. 
In such galaxies, supernova remnants in star-forming regions are a plausible source of cosmic-rays. 
The presence of cosmic-rays in galactic winds from star-forming galaxies is observationally supported by detections of high-energy gamma-ray photons from starburst galaxies, such as, M82 and NGC 253 \citep{2009Sci...326.1080A,2009Natur.462..770V,2010ApJ...709L.152A} and the Galactic diffuse soft X-ray emission\citep[e.g.,][]{1997ApJ...485..125S}. 
These emission are a probe of interactions between cosmic-rays and the ISM or radiation fields in these galaxies. 
Many theoretical models of the cosmic-ray driven wind have been developed to account for the spatial distribution and the spectral properties of the Galactic diffuse X-ray emission \citep[e.g.,][]{1991A&A...245...79B,2002A&A...385..216B,2008ApJ...674..258E}.

Although cosmic-rays could play important roles in such astrophysical situations, the development of the KH instability in the presence of cosmic-rays has not been paid attention. 
The linear analysis of the KH instability in magnetized fluid was done by \cite{1961hhs..book.....C} for the first time. 
In this work, the infinitesimal thickness of the sheared layer is assumed. 
Later, in \cite{1982JGR....87.7431M} (referred to as MP82, hereafter), the magnetized KH instability for a flow with sheared velocity profile with finite thickness was discussed. 
Then, in this paper, as a first step towards the understanding of effects of cosmic-rays on the KH instability, we extend the linear analysis of MHD equations in the presence of a sheared velocity field investigated by MP82
In other words, we perform a linear analysis of MHD equations with comic-ray effects in the same situation. 
In Section 2, our method to calculate growth rates of the KH instability is described in detail. 
We show results of the linear analysis in Section 3. 
Section 4 is devoted to summarize the results and explain how cosmic-ray pressure and diffusion affect the linear growth of the KH instability. 
Finally, Section 5 concludes this paper. 
In the following, physical variables with dimensions are denoted by letters with tilde, $\tilde{A}$, and the dimensionless counterparts are denoted by letters without tilde, $A$. 

\section{FORMULATION\label{method}}
Before moving on the linear analysis of the KH instability with cosmic-rays, we briefly review the mechanism of the KH instability in pure hydrodynamics in \S 2.1, which is helpful to understand effects of cosmic-rays. 
Next, we describe our method to deal with effects of cosmic-rays on the evolution of the hydrodynamical variables.

\subsection{KH instability in hydrodynamics and magnetohydrodynamics}
\begin{figure*}[tbp]
\begin{center}
\includegraphics[scale=0.8]{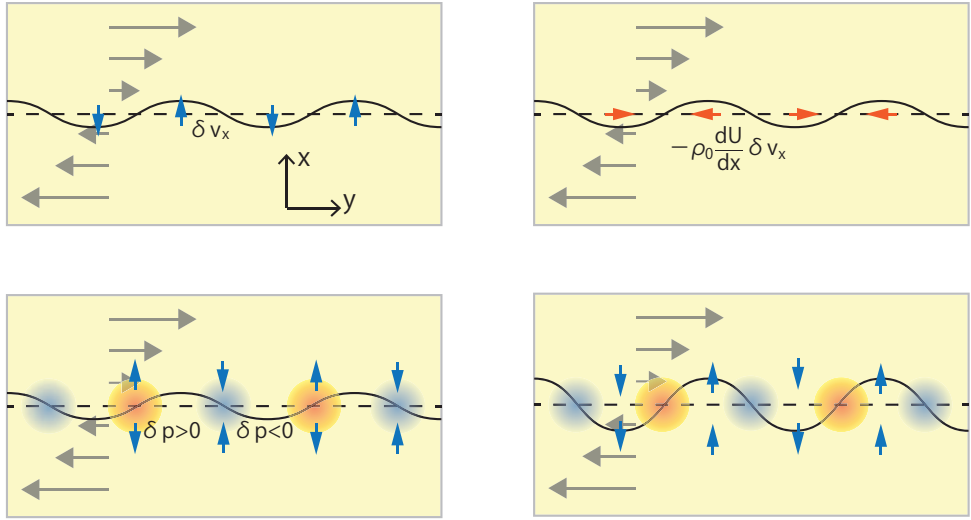}
\caption{Schematic view of the linear growth of KH instability in a plasma with a finite temperature}
\label{fig1}
\end{center}
\end{figure*}

The linear growth of the KH instability developing in a compressible fluid with a finite temperature is understood as follows. 
See the schematic view shown in Figure \ref{fig1} for the coordinates adopted here. 
i) As in the top-left panel of Figure \ref{fig1}, at the contact surface of two fluids streaming into $+y$- and $-y$-direction, a perturbation in the velocity $\delta \tilde{v}_x$ is present along the $y$-axis. 
ii) A force originated from the gradient of the velocity of the background flow acts upon fluid elements located in regions with positive (negative) values of the velocity perturbation into $-y$- ($+y$-) direction (top-right panel). 
iii) Due to the acceleration by the force, the density and the pressure at the nodes of the perturbation increases or decreases, which results in the pressure gradient and thus accelerates fluid elements around the high- and low-pressure region (bottom-left panel). 
iv) These fluid elements with increased and decreased momenta are transported along the background flow. The perturbation of the velocity is thus further enhanced by the positive feedback. 

When magnetic fields perpendicular to the direction of the background flow penetrate the system, magnetic fields serve as an additional component to the pressure of the medium. 
As a result, the condition for the instability is slightly modified. 
On the other hand, when magnetic fields parallel to the direction of the background flow exist, the magnetic tension prevents the medium from moving along the perpendicular direction of the flow, which can make the system stable.

\subsection{Basic equations of MHD with cosmic-ray}

We write down equations governing the dynamical evolution of thermal gas pressure, magnetic fields, and cosmic-ray gas pressure. 
Physical variables are functions of the coordinates $(\tilde{x},\tilde{y},\tilde{z})$ and the time $\tilde{t}$. 
The following equations describe the evolution of the gas density $\tilde{\rho}$, the velocity $\tilde{\textbf{\textit{v}}}$, the magnetic field $\tilde{\textbf{\textit{B}}}$, and the gas pressure $\tilde{p}$,
\begin{eqnarray}
\frac{\partial \tilde{\rho}}{\partial \tilde{t}}+\tilde{\nabla}\cdot(\tilde{\rho}\tilde{\textbf{\textit{v}}})&=&0,
\label{eq:continuity}\\
\rho\frac{\partial \tilde{\textbf{\textit{v}}}}{\partial \tilde{t}}+\rho(\tilde{\textbf{\textit{v}}}\cdot\tilde{\nabla})\tilde{\textbf{\textit{v}}}&=&-\tilde{\nabla}\left(\tilde{p}+\frac{\tilde{B}^2}{8\pi}+\tilde{p}_\mathrm{cr}\right)
\nonumber\\
&&\hspace{5em}+\frac{1}{4\pi}(\tilde{\textbf{\textit{B}}}\cdot\tilde{\nabla})\tilde{\textbf{\textit{B}}},
\label{eq:euler}\\
\frac{\partial \tilde{\textbf{\textit{B}}}}{\partial \tilde{t}}&=&\tilde{\nabla}\times(\tilde{\textbf{\textit{v}}}\times\tilde{\textbf{\textit{B}}}),
\label{eq:induction}\\
\frac{\partial \tilde{p}}{\partial \tilde{t}}+\tilde{\textbf{\textit{v}}}\cdot\tilde{\nabla} \tilde{p}&=&-\gamma \tilde{p}\tilde{\nabla}\cdot\tilde{\textbf{\textit{v}}},
\label{eq:energy}
\end{eqnarray}
where $\gamma(=5/3)$ is the adiabatic index and $\tilde{p}_\mathrm{cr}$ is the pressure of the cosmic-ray gas. 
Here, the density, the magnetic field, and the gas pressure obey the ideal MHD equations, the continuity equation (\ref{eq:continuity}), the induction equation (\ref{eq:induction}), and the energy equation (\ref{eq:energy}). 
The Euler equation governing the evolution of the velocity field $\tilde{\textbf{\textit{v}}}$ includes the gradient of the cosmic-ray pressure $\tilde{p}_\mathrm{cr}$ in the right-hand side of Equation (\ref{eq:euler}). 
This term expresses the feedback from the cosmic-ray gas on the thermal gas. 
The evolution of the cosmic-ray pressure introduced here is described by the following equation,
\begin{equation}
\frac{\partial \tilde{p}_\mathrm{cr}}{\partial \tilde{t}}+{\tilde\textbf{\textit{v}}}\cdot\tilde{\nabla} \tilde{p}_\mathrm{cr}=
-\gamma_\mathrm{cr}\tilde{p}_\mathrm{cr}\tilde{\nabla}\cdot\tilde{\textbf{\textit{v}}}+\tilde{\nabla}\cdot
\left[\tilde{\kappa}{\textbf{\textit{b}}}\otimes{\textbf{\textit{b}}}\cdot(\tilde{\nabla} \tilde{p}_\mathrm{cr}) \right],
\label{eq:crs}
\end{equation}
where $\gamma_\mathrm{cr}(=4/3)$ is the adiabatic index for the cosmic-ray gas and $\tilde{\kappa}$ is the diffusion coefficient parallel to the magnetic field line. 
The tensor ${\textbf{\textit{b}}}\otimes{\textbf{\textit{b}}}$ allows cosmic-rays diffuse only along the magnetic field line,
\begin{equation}
{\textbf{\textit{b}}}\otimes{\textbf{\textit{b}}}=\left(
\begin{array}{ccc}
b_xb_x&b_xb_y&b_xb_z\\
b_yb_x&b_yb_y&b_yb_z\\
b_zb_x&b_zb_y&b_zb_z\\
\end{array}
\right),
\label{eq:tensor}
\end{equation}
where ${\textbf{\textit{b}}}=\tilde{\textbf{\textit{B}}}/\tilde{B}$ is the unit vector in the direction of the magnetic field line. 
The diffusion of cosmic-rays across the magnetic field line is assumed to be negligible, since the diffusion coefficient perpendicular to the magnetic field line is approximately 100 times smaller than the parallel counterpart at least for Galactic cosmic-rays \citep{1999ApJ...520..204G}. 
Equation (\ref{eq:crs}) is derived from the cosmic-ray transport equation \citep{1971ApJ...170..265S,1975MNRAS.172..557S} under the assumption that cosmic-ray particles are efficiently scattered by Alfv$\acute\mathrm{e}$n waves propagating along magnetic field lines, i.e., the diffusion limit \citep[see, e.g.,][for details]{1985A&A...151..151S,1990acr..book.....B,2002cra..book.....S}. 

The diffusion coefficient $\tilde{\kappa}$ is thought to highly depend on the statistics of the assumed turbulent magnetic field that scatters cosmic-ray particles and thus makes the angular distribution of the cosmic-rays isotropic. 
The value is calculated by several authors and turns out to be of the order of $10^{28}$ cm$^2$/s \citep[e.g.,][]{1990acr..book.....B,2003ApJ...589..338R}.
In this paper, we adopt $\tilde{\kappa}=3\times 10^{28}$ cm$^2$/s as a fiducial value. 

\subsection{Linearized Equations}
We shall perform the linear analysis of the governing equations given in the previous section. 
First, we consider a mixture of thermal and cosmic-ray gases with uniform density and pressure as the unperturbed state,
\begin{equation}
\tilde{\rho}=\tilde{\rho_0},\ \ \ \tilde{p}=\tilde{p}_0,\ \ \ \tilde{p}_\mathrm{cr}=\alpha \tilde{p}_0.
\end{equation}
Here we have introduced the ratio $\alpha$ of the unperturbed cosmic-ray pressure to the gas pressure. 
This assumption is based on the adiabatic approximation for cosmic-rays, i.e., cosmic-rays frequently interact with Alfv$\acute{\mathrm{e}}$n waves. 
The gas is penetrated by the uniform magnetic field lying on the $y$-$z$ plane,
\begin{equation}
\tilde{\textbf{\textit{B}}}=\tilde{B}_{0y}{\textbf{\textit{e}}}_y+\tilde{B}_{0z}{\textbf{\textit{e}}}_z,
\end{equation}
where ${\textbf{\textit{e}}}_y$ and ${\textbf{\textit{e}}}_z$ are unit vectors along $y$- and $z$- axes, respectively. 
Furthermore, the following velocity profile is assumed,
\begin{equation}
\tilde{\textbf{\textit{v}}}=\tilde{U}(x){\textbf{\textit{e}}}_y=\frac{\tilde{U}_0}{2}\tanh\left(\tilde{x}/\tilde{a}\right){\textbf{\textit{e}}}_y,
\end{equation}
to realize a shear flow. 
The scales of the velocity $\tilde{U}_0$ and the length $\tilde{a}$ are used to normalize the governing equations. 
Then, the time $\tilde{t}$ is normalized by $\tilde{a}/\tilde{U}_0$ and the diffusion coefficient $\tilde{\kappa}$ is normalized by the product $\tilde{a}\tilde{U_0}$. 
The effect of cosmic-ray diffusion is expected to be prominent when the product $\tilde{a}\tilde{U}_0$ is comparable to the diffusion coefficient $\tilde{\kappa}$. 
For the fiducial value $\tilde{\kappa}=3\times 10^{28}$ cm$^2$/s, we set the value of the product $\tilde{a}\tilde{U}_0$ so that the normalized value is unity, i.e., $\tilde{a}\tilde{U}_0=3\times10^{28}$ cm$^2$/s. 
We introduce the sonic Mach number $M_\mathrm{a}$, the Alfv$\acute{\mathrm e}$n Mach number $M_\mathrm{a}$ defined in usual ways,
\begin{equation}
M_\mathrm{s}=\frac{\tilde{U}_0}{\tilde{c}_\mathrm{s}},\ \ \ 
M_\mathrm{a}=\frac{\tilde{U}_0}{\tilde{v}_\mathrm{a}},
\end{equation}
where $\tilde{c}_\mathrm{s}$ and $\tilde{v}_\mathrm{a}$ are the sound and the Alfv$\acute{\mathrm e}$n speeds corresponding to the unperturbed state,
\begin{equation}
\tilde{c}_\mathrm{s}=\sqrt{\frac{\gamma \tilde{p}_0}{\tilde{\rho}_0}},\ \ \ 
\tilde{v}_\mathrm{a}=\frac{\tilde{B}_0}{\sqrt{4\pi\tilde{\rho}_0}}
\end{equation}
Furthermore, we introduce a non-dimensional parameter $M_\mathrm{cr}$, which is expressed in terms of $\alpha$ as follows,
\begin{equation}
M_\mathrm{cr}=M_\mathrm{s}\sqrt{\frac{\gamma}{\gamma_\mathrm{cr}\alpha}}
\end{equation}
Thus, without cosmic-ray pressure, $\alpha=0$, the parameter $M_\mathrm{cr}$ diverges to $\infty$.

Next, a small perturbation, whose dependence on $t$, $y$, and $z$ are assumed to be proportional to $\exp[i(\omega t-k_yy-k_zz)]$, is added to each unperturbed physical variable. 
Substitution of the physical variables with perturbed and unperturbed parts into Equations (\ref{eq:continuity})-(\ref{eq:tensor}) leads to the following equations for the perturbations $\delta \rho$, $\delta{\textbf{\textit{v}}}$, $\delta{\textbf{\textit{B}}}$ $\delta p$, and $\delta p_\mathrm{cr}$, 
\begin{eqnarray}
\left(\omega-k_yU\right)\delta\rho&=&
-i\nabla\cdot\delta{\textbf{\textit{v}}},
\label{eq:continuity2}\\
(\omega-k_yU)\delta v_x&=&
-i\frac{d\delta p_\mathrm{tot}}{dx}
-\frac{({\textbf{\textit{k}}}\cdot{\textbf{\textit{b}}})\delta B_x}{M_\mathrm{a}^2},
\label{eq:vx}\\
(\omega-k_yU)\delta v_y&=&
-i\frac{dU}{dx}\delta v_x
+k_y\delta p_\mathrm{tot}
\nonumber\\
&&\hspace{5em}
-\frac{({\textbf{\textit{k}}}\cdot{\textbf{\textit{b}}})\delta B_y}{M_\mathrm{a}^2},
\label{eq:vy}\\
(\omega-k_yU)\delta v_z&=&
k_z\delta p_\mathrm{tot}
-\frac{({\textbf{\textit{k}}}\cdot{\textbf{\textit{b}}})\delta B_z}{M_\mathrm{a}^2},
\label{eq:vz}\\
(\omega-k_yU)\delta B_x&=&
-({\textbf{\textit{k}}}\cdot{\textbf{\textit{b}}})\delta v_x,
\label{eq:bx}\\
(\omega-k_yU)\delta B_y&=&
i\frac{dU}{dx}\delta B_x-({\textbf{\textit{k}}}\cdot{\textbf{\textit{b}}})\delta v_y
\nonumber\\
&&\hspace{5em}
-ib_{y}\nabla\cdot\delta{\textbf{\textit{v}}},
\label{eq:by}\\
(\omega-k_yU)\delta B_z&=&
-({\textbf{\textit{k}}}\cdot{\textbf{\textit{b}}})\delta v_z-ib_{z}\nabla\cdot\delta{\textbf{\textit{v}}},
\label{eq:bz}\\
(\omega-k_yU)\delta p&=&
-i\nabla\cdot\delta{\textbf{\textit{v}}},
\label{eq:p}\\
(\omega-k_yU)\delta p_\mathrm{cr}&=&
-i\nabla\cdot\delta{\textbf{\textit{v}}}
+i\kappa({\textbf{\textit{k}}}\cdot{\textbf{\textit{b}}})^2\delta p_\mathrm{cr},
\label{eq:crs2}
\end{eqnarray}
where these equations have been normalized by the length $a$, the velocity $U_0$, and some other unperturbed physical variables. 
The variables $\delta p_\mathrm{tot}$ and $\nabla\cdot\delta{\textbf{\textit{v}}}$ are a perturbation to the total pressure,
\begin{equation}
\delta p_\mathrm{tot}=
\frac{\delta p}{M_\mathrm{s}^2}+\frac{ {\textbf{\textit{b}}}\cdot\delta{\textbf{\textit{b}}}}{M_\mathrm{a}^2}+\frac{\delta p_\mathrm{cr}}{M_\mathrm{cr}^2},
\label{eq:p_ast}
\end{equation}
and the divergence of a perturbation of the velocity,
\begin{equation}
\nabla\cdot\delta{\textbf{\textit{v}}}=\left(\frac{d\delta v_x}{dx}+ik_y\delta v_y+ik_z\delta v_z\right),
\end{equation}
respectively. 
In the limit of no cosmic-ray, $M_\mathrm{cr}\rightarrow\infty$, Equations (\ref{eq:continuity2})-(\ref{eq:crs2}) reduce to the ideal MHD case. 

It is straightforward to transform Equations (\ref{eq:continuity2})-(\ref{eq:crs2}) into an ordinary differential equation for $\delta p_\mathrm{tot}$. 
According to the transformation given in Appendix, one obtains the following equation,
\begin{equation}
Q\frac{d}{dx}\left(Q^{-1}\frac{d\delta p_\mathrm{tot}}{dx}\right)+\left[\frac{(\omega-k_yU)^4}{R}-(k_y^2+k_z^2)\right]\delta p_\mathrm{tot}=0,
\label{eq:ode}
\end{equation}
where the expressions of the functions $Q$ and $R$ are given in Equations (\ref{eq:Q}) and (\ref{eq:R}). 

\subsection{Diffusion and free-streaming limits}
In the derived equation (\ref{eq:ode}), effects of cosmic-rays appear in the function $R$ that includes the two parameters $\alpha$ (or, equivalently, $M_\mathrm{cr}$) and $\kappa$. 

In the limit of negligible contribution of the cosmic-ray gas, $\alpha=0$, the function $R$ reduces to
\begin{equation}
R=\left(\frac{1}{M_\mathrm{s}^2}+\frac{1}{M_\mathrm{a}^2}\right)(\omega-k_yU)^2-
\frac{({\textbf{\textit{k}}}\cdot{\textbf{\textit{b}}})^2}{M_\mathrm{s}^2M_\mathrm{a}^2}.
\label{eq:R_zero_cr}
\end{equation}
In this limit, Equation (\ref{eq:ode}) is exactly same as the corresponding equation in the ideal MHD case in MP82. 
We can also obtain this expression by taking another limit $\kappa=\infty$, where cosmic-rays instantaneously diffuse and a uniform distribution is realized. 
In this case, the thermal and cosmic-ray gases are completely decoupled and that is why the function $R$ reduces to the ideal MHD case. 

In the opposite limit $\kappa=0$, the function $R$ approaches to the following form,
\begin{eqnarray}
R&=&\left(\frac{1}{M_\mathrm{s}^2}+\frac{1}{M_\mathrm{a}^2}+\frac{1}{M_\mathrm{cr}^2}\right)(\omega-k_yU)^2
\nonumber\\
&&\hspace{5em}
-\frac{({\textbf{\textit{k}}}\cdot{\textbf{\textit{b}}})^2}{M_\mathrm{a}^2}
\left(\frac{1}{M_\mathrm{s}^2}+\frac{1}{M_\mathrm{cr}^2}\right).
\label{eq:Rapprox}
\end{eqnarray}
This expression can be obtained by replacing $M_\mathrm{s}^{-2}$ in Equation (\ref{eq:R_zero_cr}) with $M_\mathrm{s}^{-2}+M_\mathrm{cr}^{-2}$. 
In this case, the thermal and cosmic-ray gas are completely coupled and the sound speed increases due to the contribution of cosmic-ray pressure. 

\subsection{Conditions for the Instability}
In this section, conditions for the instability to develop are discussed from the derived equation. 
An analysis similar to the ideal MHD case in MP82 can be done. 

First, we consider the incompressible limit of Equation (\ref{eq:ode}), i.e., the sonic, Alfv$\acute{\mathrm e}$n, and cosmic-ray Mach numbers, $M_\mathrm{s}$, $M_\mathrm{a}$ and $M_\mathrm{cr}$, approach to zero, 
\begin{equation}
Q\frac{d}{dx}\left(Q^{-1}\frac{d\delta p_\mathrm{tot}}{dx}\right)-(k_y^2+k_z^2)\delta p_\mathrm{tot}=0
\label{eq:ode2}
\end{equation}

From Equations (\ref{eq:vx}) and (\ref{eq:bx}), we express $\delta B_x$ in terms of the derivative of the perturbation on the total pressure $\delta p_\mathrm{tot}$,
\begin{equation}
Q\delta B_x=i({\textbf{\textit{k}}}\cdot{\textbf{\textit{b}}})\frac{d\delta p_\mathrm{tot}}{dx}
\end{equation}
Using the expression, differentiation of both sides of Equation (\ref{eq:ode2}) with respective to $x$ can be transformed into the following ordinary differential equation for $\delta B_x$,
\begin{equation}
\frac{d}{dx}\left(Q\frac{d\delta B_x}{dx}\right)-(k_y^2+k_z^2)Q\delta B_x=0
\end{equation}
Integration of this equation multiplied by the complex conjugate $\delta B_x^\ast$ from $x=-\infty$ to $x=\infty$ leads to the quadratic equation,
\begin{equation}
A\omega^2+2B\omega+C=0,
\end{equation}
where the coefficients $A$, $B$, and $C$ are given by
\begin{equation}
A=\int^\infty_{-\infty}\left[\bigg|\frac{d\delta B_x}{dx}\bigg|^2+(k_y^2+k_z^2)|\delta B_x|^2\right]dx,
\end{equation}
\begin{equation}
B=-\int^\infty_{-\infty}k_yU\left[\bigg|\frac{d\delta B_x}{dx}\bigg|^2+(k_y^2+k_z^2)|\delta B_x|^2\right]dx,
\end{equation}
and
\begin{equation}
C=\int^\infty_{-\infty}\left[k_y^2-\frac{({\textbf{\textit{k}}}\cdot{\textbf{\textit{b}}})^2}{M_\mathrm{a}^2}\right]
\left[\bigg|\frac{d\delta B_x}{dx}\bigg|^2+(k_y^2+k_z^2)|\delta B_x|^2\right]dx.
\end{equation}
It is clear that the coefficient $A$ is alway positive. 
Thus, a real $\omega$ satisfying the quadratic equation is found, if the condition $C\leq 0$ holds, which is equivalent to,
\begin{equation}
k_yM_\mathrm{a}\leq {\textbf{\textit{k}}}\cdot{\textbf{\textit{b}}}.
\end{equation}
This is exactly same as the condition derived in MP82. 
The physical interpretation of this condition is straightforward. 
A shear flow penetrated by a sufficiently strong magnetic field satisfying the above condition is stable because of the magnetic tension.

Next, we concern the behavior of the perturbation at $x=\pm \infty$, where the velocity approaches to a constant value $U(\pm \infty)=1/2$. 
Then, Equation (\ref{eq:ode}) reduces to
\begin{equation}
\frac{d^2\delta p_\mathrm{tot}}{dx^2}+\left[\frac{(\omega-k_yU)^4}{R}-(k_y^2+k_z^2)\right]
\delta p_\mathrm{tot}=0.
\end{equation}
Since any physically meaningful solution satisfies $\delta p_\mathrm{tot}\rightarrow0$ at the limits $x\rightarrow\pm\infty$, the following condition, under which the solution is an evanescent one, are required,
\begin{equation}
\mathrm{Re}\left[\frac{(\omega-k_yU)^4}{R}-(k_y^2+k_z^2)\right]<0.
\label{eq:condition}
\end{equation}
This is the condition for a physically acceptable evanescent mode to exist. 

In the following, we assume that the frequency $\omega$ is much smaller than $k_yU$, i.e., the short wavelength limit, and we again take $\kappa=0$, under which the function $R$ reduces to Equation (\ref{eq:Rapprox}). 
In order to clarify effects of cosmic-rays, we consider a special case ${\textbf{\textit{k}}}\cdot{\textbf{\textit{b}}}=0$. 
The condition (\ref{eq:condition}) can be expressed as
\begin{equation}
\frac{k_y^2}{4(k_y^2+k_z^2)}<\frac{1}{M_\mathrm{s}^2}+\frac{1}{M_\mathrm{a}^2}+\frac{1}{M_\mathrm{cr}^2},
\label{eq:condition_perp}
\end{equation}
after replacing the function $U$ with the asymptotic value $U(\pm\infty)=1/2$. 
The same expression can be obtained by replacing $M_\mathrm{s}^{-2}$ in the corresponding condition for the ideal MHD case with $M_\mathrm{s}^{-2}+M_\mathrm{cr}^{-2}$. 
In the ideal MHD case, the velocity of a flow satisfying the above condition needs to be slower than the magnetosonic speed for an evanescent mode to exist. 
In the presence of cosmic-rays, we can see that the condition is relaxed due to the additional term, $M_\mathrm{cr}^{-2}$. 
It is because cosmic-ray pressure increases the effective magnetosonic speed and thus subsonic flows are easy to realize in comparison with the ideal MHD case.

\section{EIGEN-VALUE PROBLEM}

\begin{figure}[tbp]
\begin{center}
\includegraphics[scale=0.45]{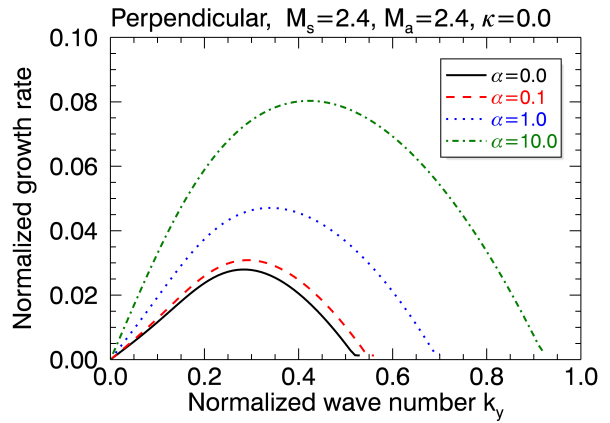}
\caption{Growth rates as functions of the wave number $k_y$ for the perpendicular case with $M_\mathrm{s}=2.4$, $M_\mathrm{a}=2.4$, $\kappa=0$, $k_z=0$ and different $\alpha$.}
\label{fig2}
\end{center}
\end{figure}

\begin{figure*}[tbp]
\begin{center}
\includegraphics[scale=0.8]{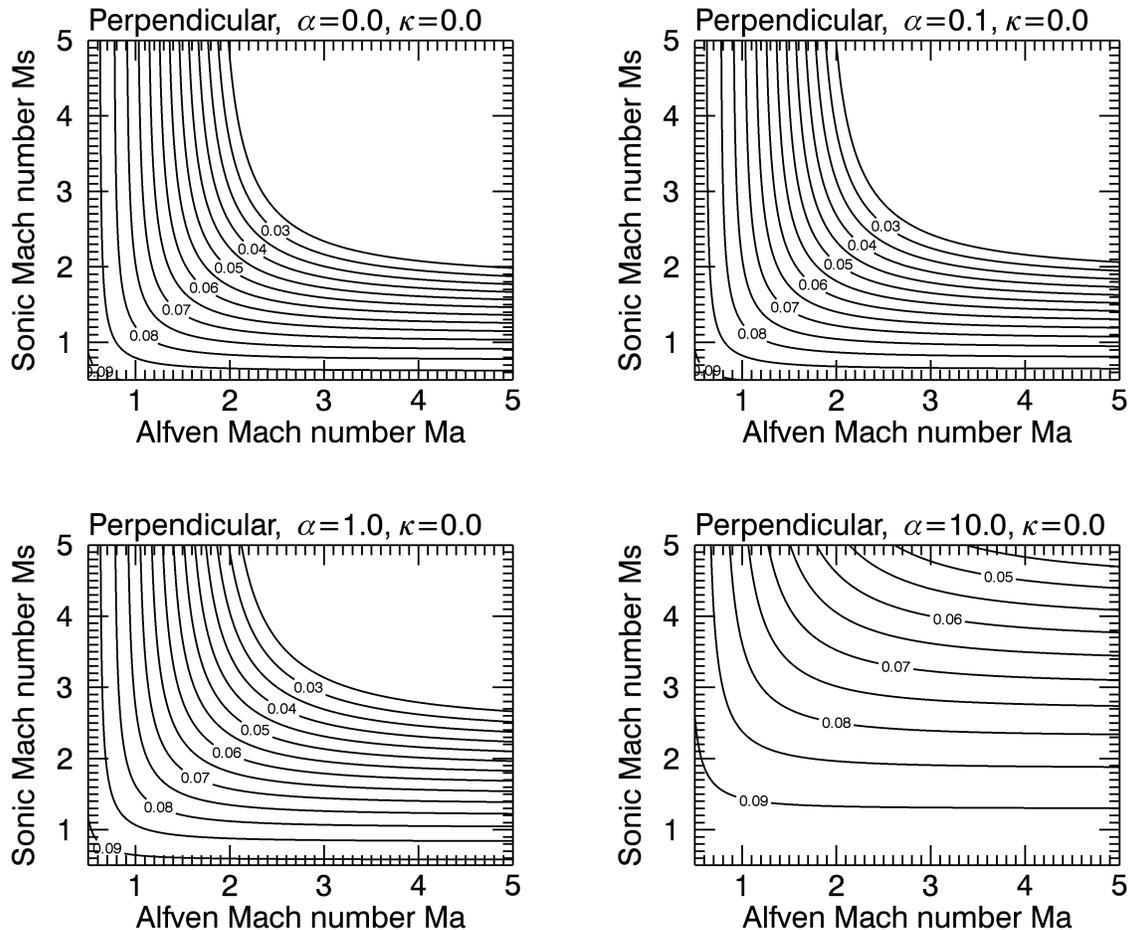}
\caption{Maximum growth rate as a function of the sonic and Alfv$\acute{\mathrm{e}}$n Mach number $M_\mathrm{s}$ and $M_\mathrm{a}$ for a perpendicular magnetic field. The parameters are set to $\alpha=0.0$, $0.1$, $1.0$, and $10.0$, and $\kappa=0.0$.}
\label{fig:perp}
\end{center}
\end{figure*}

Now we solve the derived ordinary differential equation (\ref{eq:ode}) under appropriate boundary conditions. 
In this section, we describe our method to calculate the growth rate of the instability and the corresponding eigen function and show results of the calculation. 

\subsection{Boundary Conditions}
In the ordinary differential equation (\ref{eq:ode}), the real part of the frequency $\omega$ is assumed to be zero and we seek the imaginary part $\omega_\mathrm{i}$ that satisfies the boundary condition. 
In other words, the frequency $i\omega_\mathrm{i}$ is determined as an eigen value. 
We integrate Equation (\ref{eq:ode}) from the outer boundary $x=x_\mathrm{out}$ to the inner boundary $x=0$. 
It is necessary to locate the outer boundary far from $x=0$ so that its influence on the solution is negligible. 
We adopt $x_\mathrm{out}=20$ in this study. 
We have tried calculations with $x_\mathrm{out}=30$ and confirmed that results are not different from those with $x_\mathrm{out}=20$. 
Thus, we present results of calculations with $x_\mathrm{out}=20$ in the following. 
The integration is carried out by using the 4th-order Runge-Kutta scheme. 

At the outer boundary $x=x_\mathrm{out}$, the differentiation of the perturbed total pressure is set to zero, $d\delta p_\mathrm{tot}/dx=0$. 
As the equation has been linearized, the amplitude of $\delta p_\mathrm{tot}$ is arbitrary. 
On the other hand, the phase $\theta$ of $\delta p_\mathrm{tot}$ is unknown. 
Therefore, we set $\delta p_\mathrm{tot}=e^{i\theta}$ at $x=x_\mathrm{out}$. 
The phase $\theta$ is also an eigen value to be determined. 
Because of the symmetry of the velocity profile $U=\tanh(x)/2$, it turns out that the real part of $\delta p_\mathrm{tot}$ and the imaginary part of  $d\delta p_\mathrm{tot}/dx$ are zero at the inner boundary $x=0$. 

In summary, we integrate the ordinary differential equation from the outer boundary, where the following condition holds,
\begin{equation}
\delta p_\mathrm{tot}=e^{i\theta},\ \ \ \frac{d\delta p_\mathrm{tot}}{dx}=0,
\end{equation}
to the inner boundary for a given set of parameters $k_y,k_z,\kappa,M_\mathrm{s},M_\mathrm{a},$ and $M_\mathrm{cr}$. 
The equation and the initial condition include two unknown parameters $\omega_\mathrm{i}$ and $\theta$. 
These values are determined to satisfy the condition at the inner boundary,
\begin{equation}
\mathrm{Re}[\delta p_\mathrm{tot}]=0,\ \ \ 
\mathrm{Im}\left[\frac{d\delta p_\mathrm{tot}}{dx}\right]=0.
\end{equation}
The root-finding algorithm we employ is the standard Newton-Raphson method. 

To check that the method works well, we have calculated the dispersion relation without cosmic-rays $\alpha=0$, i.e., the ideal MHD case. 
The same problem has been solved by MP82, where a different method to calculate growth rates of the instability is employed. 
We confirmed that our results for the ideal MHD case are in good agreement with theirs.

\subsection{Dispersion Relations\label{result}}
We present growth rates of the KH instability as functions of the wave number for a few sets of parameters. 
Especially, the orientation of the magnetic field is of great importance. 
In the following, we consider magnetic fields perpendicular and parallel to $y$-axis. 

\subsubsection{Perpendicular case}
At first, we show growth rates of the KH instability as functions of $k_y$ for $\alpha=0,0.1,1.0,$ and $10$ and $\kappa=0$ for a perpendicular magnetic field in Figure \ref{fig2}. 
Thus, ${\textbf{\textit{k}}}\cdot{\textbf{\textit{b}}}=0$ holds. 
The $z$-component of the wave vector is fixed to be zero and the other parameters are set to be $M_\mathrm{s}=M_\mathrm{a}=2.4$. 
The curve labeled by $\alpha=0$ corresponds to the ideal MHD case. 
It turns out that larger $\alpha$ result in larger growth rates. 
This clearly demonstrates the cosmic-ray pressure enhances the development of the KH instability. 
Figure \ref{fig:perp} shows the maximum growth rates as functions of $M_\mathrm{s}$ and $M_\mathrm{a}$ for $\alpha=0.0$, $0.1$, $1.0$, and $10.0$. 
While the instability does not develop for large $M_\mathrm{s}$ and $M_\mathrm{a}$ without cosmic-rays (top-left panel), the system becomes unstable even for large $M_\mathrm{s}$ and $M_\mathrm{a}$  when the cosmic-ray pressure dominates over the thermal pressure (bottom-right panel). 

We also performed calculations of growth rates for finite diffusion coefficients $\kappa\neq 0$. 
However, in this configuration of the magnetic field, growth rates do not depend on the value of $\kappa$. 
The reason is discussed in the next section. 

\subsubsection{Parallel case}
\begin{figure}[tbp]
\begin{center}
\includegraphics[scale=0.45]{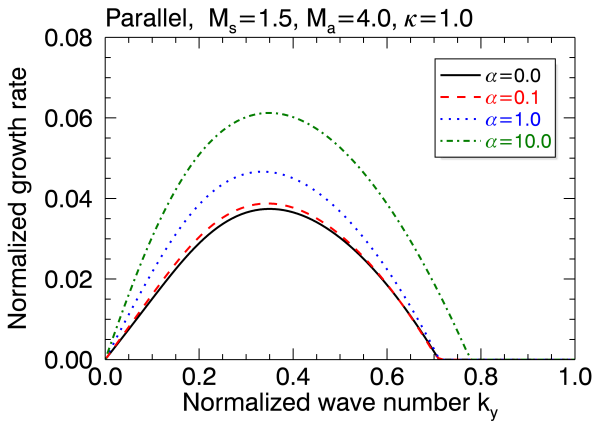}
\caption{Growth rates as functions of the wave number $k_y$ for the parallel case with $M_\mathrm{s}=1.5$, $M_\mathrm{a}=4.0$, $\kappa=0$, $k_z=0$ and different $\alpha$.}
\label{fig3}
\end{center}
\end{figure}

Figure \ref{fig3} shows growth rates for $\alpha=0,0.1,1.0,$ and $10$ and $\kappa=0$ for a parallel magnetic field. 
The behavior of the growth rates for increasing $\alpha$ is same as the perpendicular case. 

\begin{figure}[tbp]
\begin{center}
\includegraphics[scale=0.45]{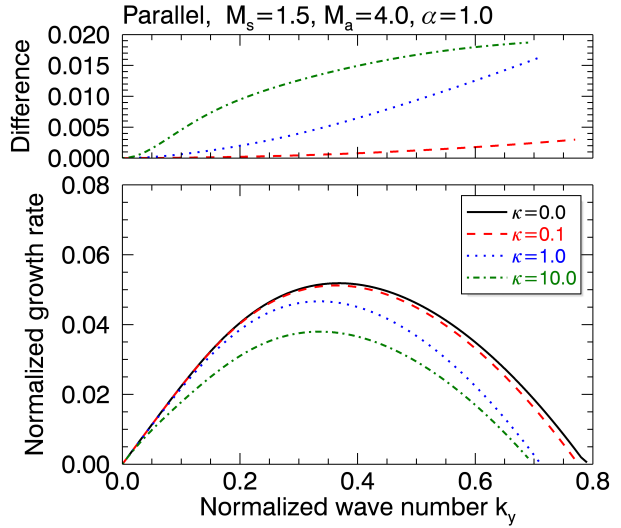}
\caption{Growth rate as a function of the wave number $k_y$ for the parallel case with $M_\mathrm{s}=1.5$, $M_\mathrm{a}=4.0$, $\alpha=1.0$, $k_z=0$ and different $\kappa$ (the lower panel) and the absolute value of the deviation of each growth rate from the case with no diffusion $\kappa=0$ (the upper panel).}
\label{fig4}
\end{center}
\end{figure}

The dependence of the growth rates for $\alpha=1.0$ and $\kappa=0.1,1.0,10$ is shown in Figure \ref{fig4}. 
The ideal MHD case is also plotted for the purpose of comparison. 
In the presence of a parallel magnetic field, contrary to a perpendicular one, growth rates of the instability depend on the diffusion coefficient. 
Basically, larger values of the diffusion coefficient result in smaller growth rates. 
The curve for the largest $\kappa(=10)$ in Figure \ref{fig4} is almost identical with that of the ideal MHD case. 
In other words, the diffusion suppresses the enhancement of the growth rate due to cosmic-ray pressure. 
As seen in the curve for $\kappa=1.0$ in the lower panel of Figure \ref{fig4}, the suppression is more significant for larger $k_y$, i.e., the short-wavelength regime. 

\begin{figure*}[tbp]
\begin{center}
\includegraphics[scale=0.8]{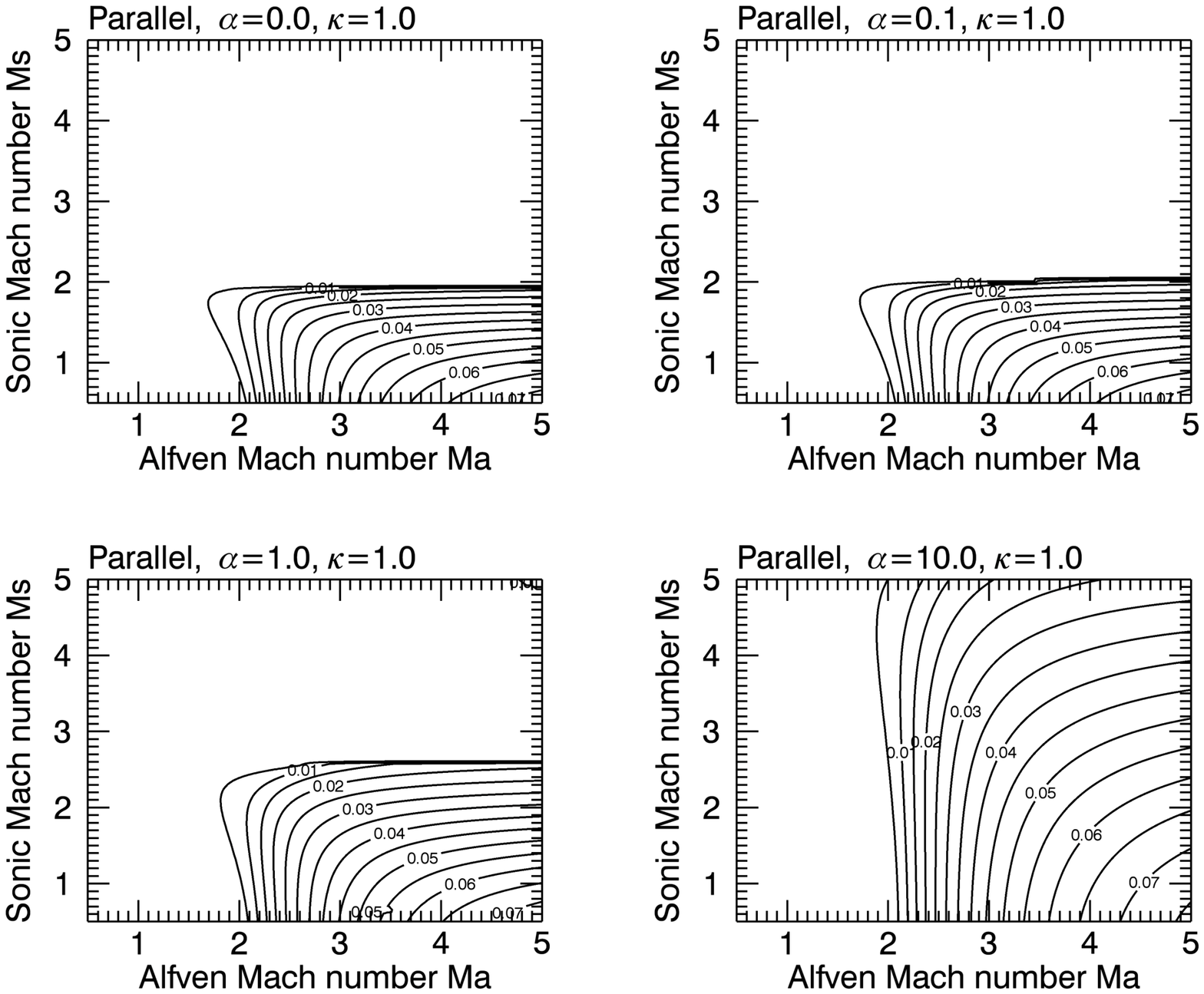}
\caption{Maximum growth rate as a function of the sonic and Alfv$\acute{\mathrm{e}}$n Mach number $M_\mathrm{s}$ and $M_\mathrm{a}$ for a parallel magnetic field. The parameters are set to $\alpha=0.0$, $0.1$, $1.0$, and $10.0$, and $\kappa=1.0$.}
\label{fig:para_b}
\end{center}
\end{figure*}

\begin{figure*}[tbp]
\begin{center}
\includegraphics[scale=0.8]{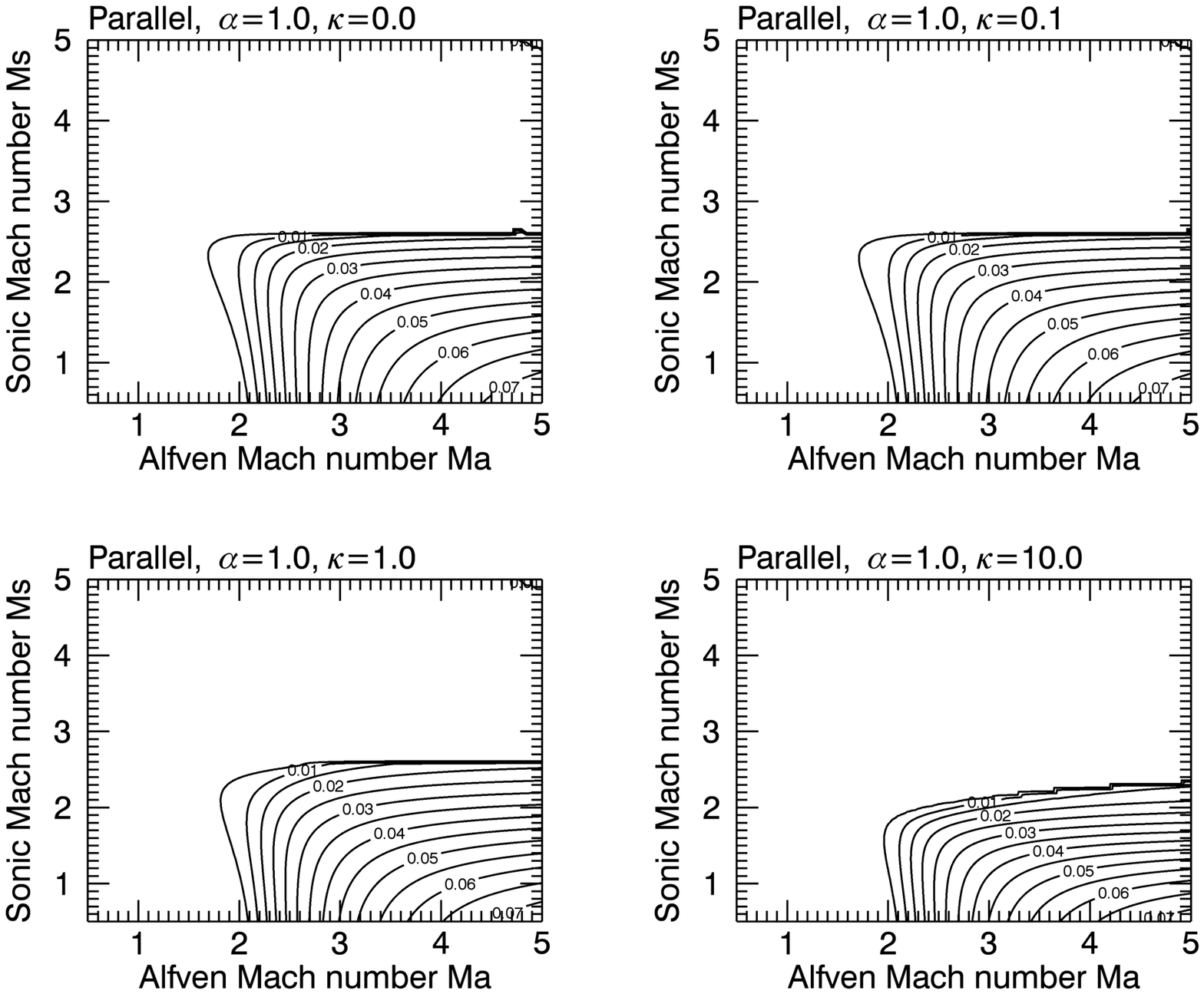}
\caption{Maximum growth rate as a function of the sonic and Alfv$\acute{\mathrm{e}}$n Mach number $M_\mathrm{s}$ and $M_\mathrm{a}$ for a parallel magnetic field. The parameters are set to $\alpha=0.0$ and $\kappa=0.0$, $0.1$, and $1.0$}
\label{fig:para_k}
\end{center}
\end{figure*}

The maximum growth rates as functions of $M_\mathrm{s}$ and $M_\mathrm{a}$ for different $\alpha$ and $\kappa$ are shown in Figures \ref{fig:para_b} and \ref{fig:para_k}. 
In cases with parallel magnetic fields, the KH instability cannot develop for small $M_\mathrm{a}$. 
This is because flows along the perpendicular direction of sufficiently strong magnetic fields are not allowed due to the magnetic tension. 
The dependence on $\alpha$ is similar to the perpendicular case, i.e., the system can be unstable for large $M_\mathrm{s}$ and $M_\mathrm{a}$ when the cosmic-ray pressure dominates over the thermal pressure. 
On the other hand, the unstable region on the $M_\mathrm{s}$-$M_\mathrm{a}$ plot becomes smaller when larger values of $\kappa$ are adopted. 

\subsection{Eigen functions}
\begin{figure}[tbp]
\begin{center}
\includegraphics[scale=0.45]{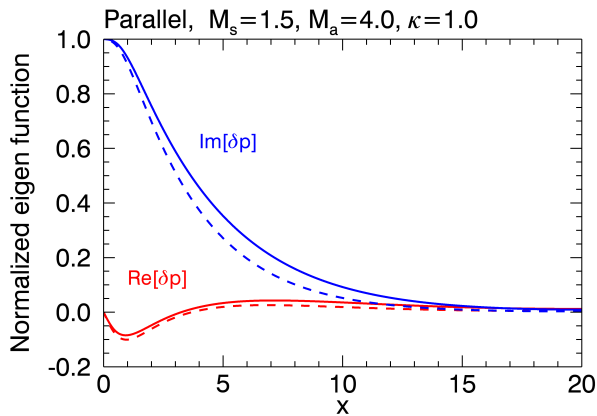}
\caption{Real and imaginary parts of eigen functions normalized so that the imaginary part is unity at $x=0$. Solid and dashed lines correspond to calculations with $\alpha=1$ and $\alpha=0$. Other parameters are set to be $M_\mathrm{s}=1.5$, $M_\mathrm{a}=4.0$, $k_y=0.4$, $k_z=0$, and $\kappa=1.0$.}
\label{fig5}
\end{center}
\end{figure}

For each set of eigen values, we can calculate the eigen function $\delta p_\mathrm{tot}$. 
Figure \ref{fig5} shows the profile of the eigen functions for the parallel case with $k_y=0.4$ and $k_z=0.0$. 
Other parameters are set to $M_\mathrm{s}=1.5$, $M_\mathrm{a}=4.0$, and $\alpha=1.0$. 
Solid and dashed curves correspond to the models with $\kappa=0$ and $\kappa=1$. 
Actually, the profiles of the function are similar to the ideal MHD case even when cosmic-ray diffusion is included.

\section{COSMIC-RAY EFFECTS}
In this section, we summarize the result of the linear analysis and discuss how cosmic-rays behave in the linearly growing phase of the KH instability. 

\subsection{Effect of Cosmic-ray Pressure}
The effect of the cosmic-ray pressure is governed by the ratio of the cosmic-ray pressure to the thermal gas pressure, $\alpha$.
As we have shown in Section \ref{result}, the increase in the parameter $\alpha$ enhances the growth of the instability in both perpendicular and parallel configurations of the magnetic field. 
The physical interpretation of this effect is straightforward. 
In the linear regime, the KH instability develops because the regions where positive and negative pressure gradient perpendicular to the direction of the flow emerge along the interface (see the bottom left panel of Figure \ref{fig1}). 
The strength of the pressure gradient reflects the total pressure perturbation, which is composed of thermal gas, magnetic, and cosmic-ray pressures. 
Therefore, the increase in the cosmic-ray pressure leads to a steep pressure gradient. 
This explains the tendency of the growth rate when we assume larger $\alpha$. 

For cases with perpendicular magnetic fields, the inequality derived in the previous section, Equation (\ref{eq:condition_perp}), must hold for the instability to develop. 
This means that the system is stable for large $M_\mathrm{s}$ and $M_\mathrm{a}$ (see the panels in Figure \ref{fig:perp}).  
Large values of $\alpha$ make it easy for $M_\mathrm{s}$ and $M_\mathrm{a}$ to satisfy the inequality. 
This explains the reason why the unstable region on the $M_\mathrm{s}$-$M_\mathrm{a}$ plot becomes larger when larger values of $\alpha$ are assumed.

\subsection{Effect of Cosmic-ray Diffusion}
The dependence of the growth rate on the diffusion coefficient is understood as follows. 
At first, the growth rate turns out to be independent of $\kappa$ in the perpendicular case. 
In this case, the inner product ${\textbf{\textit{k}}}\cdot{\textbf{\textit{b}}}$ is zero, which means that Equation (\ref{eq:ode}) reduces the same form as the ideal MHD case with the replacement of $M_\mathrm{s}^{-2}$ with $M_\mathrm{s}^{-2}+M_\mathrm{cr}^{-2}$. 
In other words, the cosmic-ray gas can affect the evolution of the system only through the cosmic-ray pressure. 
In the following discussion, we consider the case $k_z=0$, as is shown in Figure \ref{fig1}. 
In this case, the magnetic field lines lie along $z$-axis and thus the pressure perturbations take the same value along the magnetic field line. 
Since the cosmic-ray gas diffuses only along the magnetic field line, it is obvious that no diffusion occurs in such configuration. 

On the other hand, in the parallel case, the growth rate can decrease by diffusion, especially, in the short-wavelength regime.  
In this case, the magnetic field lines lying around $x=0$ (see the bottom left panel of Figure \ref{fig1}) connect the nodes of the perturbation $\delta v_x$, where regions with positive and negative pressure perturbations are present side by side. 
Cosmic-rays around a node with positive $\delta p$ are allowed to diffuse into the neighboring nodes, which decreases the pressure gradient of the region with positive $\delta p$. 
Sufficiently large value of the diffusion coefficient makes the spatial distribution of the cosmic-ray gas uniform and thus suppress the effect of cosmic-ray pressure. 

For moderate values of the diffusion coefficient, e.g., $\kappa=1$, the suppression appears only in the short-wavelength regime. 
In Equation (\ref{eq:R}), it is found that cosmic-rays efficiently suppress the development of the instability when the term $\kappa({\textbf{\textit{k}}}\cdot{\textbf{\textit{b}}})^2$ is larger than $\omega$ and $k_yU$. 
This can be understood as follows. 
We can define a time scale $t_\mathrm{ch}$ by dividing the characteristic length $k_y^{-1}$ of the perturbation by the asymptotic speed $U(\pm\infty)=1/2$,
\begin{equation}
t_\mathrm{ch}=\frac{2}{k_y}
\end{equation}
On the other hand, the time scale for cosmic-rays to carry energy from a $\delta p>0$ region to two neighboring $\delta p<0$ regions by diffusion, is given by the diffusion time scale,
\begin{equation}
t_\mathrm{diff}=\frac{1}{({\textbf{\textit{k}}}\cdot{\textbf{\textit{b}}})^2\kappa}.
\end{equation}
If the diffusion time scale is shorter than the characteristic time scale $t_\mathrm{ch}$ and the growth time scale $1/\omega$, there is sufficient time for diffusion to interrupt the development of the instability. 
The condition $t_\mathrm{diff}<t_\mathrm{ch}$ leads to
\begin{equation}
\kappa>\frac{k_y}{2({\textbf{\textit{k}}}\cdot{\textbf{\textit{b}}})^2}.
\end{equation}
This explains why cosmic-ray diffusion affects more on modes with shorter wavelengths for a given diffusion coefficient.

\section{SUMMARY AND DISCUSSIONS}
In this study, we investigate the linear growth of the KH instability in the presence of cosmic-ray gas. 
MHD equations incorporated with cosmic-ray pressure are linearized and then solved as an eigen-value problem to obtain growth rates in the linear phase. 
Our results show 
i) cosmic-ray pressure enhances the growth of the instability, and 
ii) cosmic-ray diffusion can suppress the enhancement in the magnetic field parallel to the flow. 
The suppression is more effective for perturbations with shorter wavelengths. 
Especially, when the cosmic-ray pressure dominates over the thermal gas pressure, the instability can develop even for large sonic and Alfv$\mathrm{\acute{e}}$n Mach numbers that stabilize the system without cosmic-ray effects. 
Therefore, cosmic-ray effects are prominent for the system with the flow velocity larger by a factor than the sound and the Aflv$\mathrm{\acute{e}}$n velocity, $M_\mathrm{s},M_\mathrm{a}>3$. 

\subsection{Values of the cosmic-ray pressure to the thermal gas pressure ratio $\alpha$}
It is found that the parameter $\alpha$, the ratio of the cosmic-ray pressure to the thermal pressure, is crucial for the development of the KH instability. 
In star-forming galaxies, the value is expected to be similar to the Galactic value, $\alpha\sim 1$. 
\cite{2008ApJ...674..258E} developed a galactic outflow model including cosmic-rays based on the model originally presented in \cite{1991A&A...245...79B}. 
Their model implies that the cosmic-ray pressure is comparable to the thermal gas pressure. 
However, assuming that supernova remnants predominantly produce Galactic cosmic-rays, cosmic-ray energy density distribution in a star-forming galaxy obeys the spatial distribution of star-forming regions in the galaxy. 
It is naturally expected that some regions with large values of $\alpha$ exist locally while the value is globally unity.  

Furthermore, the value of $\alpha$ might be larger in starburst galaxies than that in the Galaxy inferred from the model of \cite{2008ApJ...674..258E}.  
It is known that starburst galaxies, such as, M82 and NGC 253, are known to be gamma-ray sources \citep[e.g.,][]{2009Sci...326.1080A,2009Natur.462..770V,2010ApJ...709L.152A}. 
These gamma-ray emission are a tracer of cosmic-rays in these galaxies, because they are thought to be emitted by the interaction between cosmic-rays and ISM gas or radiation fields in these galaxies. 
The cosmic-ray energy density at the nuclei of starburst galaxies is estimated to be several hundreds to thousands times higher than that of the Galactic value. 
This is supported by an independent estimation of the cosmic-ray energy density at the nuclei by ultraluminous infrared galaxies \citep{2010ApJ...720..226P} from infrared observations. 

\subsection{Possible sites of KH instability with cosmic-rays}
The cosmic-ray effects considered in this paper may be important in the following astrophysical sites. 

\subsubsection{Outflows from star-forming galaxies}
In the best-fit model of \cite{2008ApJ...674..258E}, the sonic and Alfv$\acute{\mathrm{e}}$n Mach numbers of the flow increase as the height from the disk increases and reaches to a few at the height of 10  kpc. 
The KH instability would not develop in the region without cosmic-rays because the flow is supersonic with $M_\mathrm{s},M_\mathrm{a}>2$-$3$. 
However, the cosmic-ray pressure is slightly larger than the thermal gas pressure in the region, suggesting the development of the instability supported by cosmic-ray pressure. 
For the adopted value of the cosmic-ray diffusion coefficient, $\tilde{\kappa}=3\times 10^{28}$ cm$^2$ s$^{-1}$ and the terminal velocity, $\sim 800$ km s$^{-1}$, of the outflow inferred from their model, the scale length $\tilde{a}$ is about $0.1$ kpc. 
If we consider the parallel case with $\alpha=1.0$ and $\kappa=1.0$ in Figure \ref{fig:para_b}, in which the wave number that gives the maximum growth rate is found to be $k_y\sim 0.3$, perturbations with scales of $\sim 0.3$ kpc are expected to grow efficiently. 
The growing time scale $\tilde{t}_\mathrm{grow}$ is estimated to be
\begin{eqnarray}
\tilde{t}_\mathrm{grow}=\frac{\tilde{a}}{\gamma_\mathrm{max}\tilde{U}_0}
&=&2\times 10^{6}\ \mathrm{yr}\left(\frac{\tilde{a}}{0.1\ \mathrm{kpc}}\right)
\nonumber\\
&&
\hspace{-2em}\times\left(\frac{\gamma_\mathrm{max}}{0.05}\right)^{-1}
\left(\frac{\tilde{U}_0}{800\ \mathrm{km}\ \mathrm{s}^{-1}}\right)^{-1},
 \end{eqnarray}
where $\gamma_\mathrm{max}$ is the maximum growth rate. 
On the other hand, the time required for the gas moving at $800$ km s$^{-1}$ to travel $10$ kpc is about $10^7$ yr. 
Therefore, the KH instability supported by cosmic-rays can develop in the region.

\subsubsection{AGN jets}
In order to explain high-energy emission produced by jets from active galactic nuclei(AGN), stratified jet models have been put forward. 
In such models, a jet is divided into two component, the material moving at relativistic speeds and the surrounding material moving at relatively low velocities. 
The latter component is realized as a result of the interaction between the jet and the ambient medium, which can be found in relativistic hydrodynamical simulations \citep[see, e.g.,][]{1999ApJ...523L.125A}. 
In such circumstances, a cocoon filled with cosmic-rays accelerated at the boundary between the components  can form and give rise to non-thermal emission by synchrotron and inverse compton processes \citep{2000MNRAS.312..579O,2002ApJ...578..763S,2008ApJ...681.1725S}. 
The cosmic-rays are expected to accelerate via 2nd-order Fermi process in a tangled magnetic field produced by the KH instability at the boundary. 
If that is the case, the thus produced cosmic-rays enhance the development of the KH instability and the efficiency of the acceleration process. 
Recently, \cite{2011ApJ...728..121G} performed numerical simulations of the propagation of a cosmic-ray dominated AGN jet in the ICM. 
They found that the structure created in the ICM as a result of the injection of cosmic-ray dominated jet look similar to observed X-ray cavities in the ICM of some clusters of galaxies \citep[see, e.g.,][]{2007ARA&A..45..117M}, while kinetic energy dominated jets create relatively elongated cavity. 
In their calculations, cosmic-ray energy pressure is dominant over the thermal gas pressure in the inner region of the jet. 
Thus the value of $\alpha$ is considerably large in the region and cosmic-ray effects on the KH instability considered in this paper can be expected.

\subsubsection{Channel flows as a result of magnetorotational instability}
The third example is the magnetorotational instability in differentially rotating accretion disks \citep{1991ApJ...376..214B}. 
Numerical simulations of the instability have found that the instability results in the formation of channel flows \citep{1992ApJ...400..595H}. 
As the channel flows grow, the KH instability develops and then destroys the channel flow, which leads to the non-linear saturation of the magnetorotational instability \citep{1994ApJ...432..213G}. 
The KH instability would also be pronounced on the surface of the accretion disks. 
Due to the amplification of the magnetic field by MRI, the disk wind (outflow) is formed \citep{1982MNRAS.199..883B}. 
The outflow is unstable on the dynamical timescale, which is potentially due to the KH instability \citep{2013A&A...550A..61L}. 
This results in a formation of turbulent outflow. 
In the presence of cosmic-rays, the non-linear saturation process should be affected. 

\subsubsection{Galactic disks}
Another possible site for the KH instability with cosmic-rays is galactic disks, where a gas flows in a spiral potential of the galaxy. 
\cite{2004MNRAS.349..270W} have studied stability of gas flows in several spiral potentials and found that clumps are formed in shear layers where gases are compressed by the spiral shock. 
They attribute the generation of clumps to the KH instability developing in the shear layer. 
For a galaxy with the physical scale of several kpc and the velocity of the rotation of 100 km s$^{-1}$, the product of the physical scale and the velocity, which is of the order of $10^{28}$ cm$^2$ s$^{-1}$, can be comparable to the the diffusion coefficient of galactic cosmic-rays and thus cosmic-rays may affect the development of the KH instability. 

\subsection{Some remarks}

While we examined how cosmic-rays affect the linearly growing stage of the KH instability, their influence on the non-linear stage is also of great interest. 
In the non-linear stage of the instability, it is known that the instability destroys the shear flow and makes the fluid turbulent. 
Investigating effects of cosmic-rays on the creation of vortices in a turbulent fluid requires hydrodynamical simulations incorporated with cosmic-ray diffusion \citep{2004ApJ...605L..33H}. 
Such simulations are one of the future works. 
Furthermore, although we treat cosmic-rays as an adiabatic gas, it is also important to consider the KH instability in the presence of non-thermal particles with a particular energy spectrum. 
In order to tackle the problem, it is necessary to solve hydrodynamical equations with cosmic-ray transport in energy space, which is also a future work of special interests. 

\acknowledgments
We are grateful to an anonymous referee for his/her constructive comments on the manuscript. 
Numerical computations were in part carried out on the general-purpose PC farm at Center for Computational Astrophysics, National Astronomical Observatory of Japan. 
A part of this research has been funded by MEXT HPCI STRATEGIC PROGRAM. 
A.S. is supported by a Grant-in-Aid for JSPS Fellows. 
T.K. is supported by a Grant-in-Aid for Scientific Research (KAKENHI)
from JSPS (P.I. TK:23540274). 
This work was also supported in part by the Center for the Promotion of
Integrated Sciences (CPIS) of Sokendai.

\appendix
\section{\label{appendix}DERIVATION OF THE ORDINARY DIFFERENTIAL EQUATION FOR THE TOTAL PRESSURE PERTURBATION}
In this section, we briefly describe the way to reduce Equations (\ref{eq:continuity2})-(\ref{eq:crs2}) to the ordinary differential equation (\ref{eq:ode}). 

At first, eliminating $\delta B_x$ from the right-hand side of Equation (\ref{eq:vx}) using Equations (\ref{eq:bx}), one obtains the following equation,
\begin{equation}
(\omega-k_yU)^2\delta v_x=-i(\omega-k_yU)\frac{d}{dx}\delta p_\mathrm{tot}+\frac{({\textbf{\textit{k}}}\cdot{\textbf{\textit{b}}})^2}{M_\mathrm{a}^2}\delta v_x.
\label{eq:delta_x}
\end{equation}
In similar ways, eliminations of $\delta B_y$ and $\delta B_z$ from the right-hand sides of Equations (\ref{eq:vy}) and (\ref{eq:vz}) using Equations (\ref{eq:by}) and (\ref{eq:bz}) result in
\begin{equation}
(\omega-k_yU)^2\delta v_y=
-i\frac{dU}{dx}\frac{Q}{\omega-k_yU}\delta v_x
+k_y(\omega-k_yU)\delta p_\mathrm{tot}
-\frac{({\textbf{\textit{k}}}\cdot{\textbf{\textit{b}}})}{M_\mathrm{a}^2}\delta v_y
+i\frac{({\textbf{\textit{k}}}\cdot{\textbf{\textit{b}}})}{M_\mathrm{a}^2}
b_y\nabla\cdot \delta{\textbf{\textit{v}}},
\end{equation}
and 
\begin{equation}
(\omega-k_yU)^2\delta v_y=
k_z(\omega-k_yU)\delta p_\mathrm{tot}
-\frac{({\textbf{\textit{k}}}\cdot{\textbf{\textit{b}}})}{M_\mathrm{a}^2}\delta v_z
+i\frac{({\textbf{\textit{k}}}\cdot{\textbf{\textit{b}}})}{M_\mathrm{a}^2}
b_z\nabla\cdot \delta{\textbf{\textit{v}}},
\end{equation}
where we have defined the function $Q$ as,
\begin{equation}
Q=(\omega-k_yU)^2-\frac{({\textbf{\textit{k}}}\cdot{\textbf{\textit{b}}})^2}{M_\mathrm{a}^2}.
\label{eq:Q}
\end{equation}
From the derived expressions for $\delta v_x$, $\delta v_y$ and $\delta v_z$, we can derive the following expression for the divergence $\nabla\cdot\delta {\textbf{\textit{v}}}$,
\begin{equation}
(\omega-k_yU)\nabla\cdot\delta {\textbf{\textit{v}}}=
-iQ\frac{d}{dx}\left(Q^{-1}\frac{d\delta p_\mathrm{tot}}{dx}\right)
+i(k_y^2+k_z^2)\delta p_\mathrm{tot}.
\label{eq:div}
\end{equation}

On the other hand, elimination of $\delta B_x$ and $\delta v_y$ in the right-hand side of Equations (\ref{eq:by}) leads to
\begin{equation}
Q\delta B_y=
-k_y({\textbf{\textit{k}}}\cdot{\textbf{\textit{b}}})\delta p_\mathrm{tot}
-ib_y\nabla\cdot\delta{\textbf{\textit{v}}}.
\end{equation}
In a similar way, the following expression for $\delta B_z$ can be obtained,
\begin{equation}
Q\delta B_z=
-k_z({\textbf{\textit{k}}}\cdot{\textbf{\textit{b}}})\delta p_\mathrm{tot}
-ib_z\nabla\cdot\delta{\textbf{\textit{v}}}. 
\label{eq:delta_bz}
\end{equation}
From these expressions and Equations (\ref{eq:p}) and (\ref{eq:crs2}), the perturbed total pressure (\ref{eq:p_ast}) is expressed as follows,
\begin{equation}
(\omega-k_yU)^2\delta p_\mathrm{tot}=
-i\frac{R}{\omega-k_yU}\nabla\cdot\delta{\textbf{\textit{v}}},
\label{eq:p_ast2}
\end{equation}
where the function $R$ is defined as,
\begin{equation}
R=
\left[\left(\frac{1}{M_\mathrm{s}^2}+\frac{1}{M_\mathrm{a}^2}\right)(\omega-k_yU)^2-\frac{({\textbf{\textit{k}}}\cdot{\textbf{\textit{b}}})^2}{M_\mathrm{s}^2M_\mathrm{a}^2}\right]
+\frac{1}{M_\mathrm{cr}^2}\frac{\omega-k_yU}{\omega-k_yU+i\kappa({\textbf{\textit{k}}}\cdot{\textbf{\textit{b}}})^2}\left[(\omega-k_yU)^2-\frac{({\textbf{\textit{k}}}\cdot{\textbf{\textit{b}}})^2}{M_\mathrm{a}^2}\right]
\label{eq:R}
\end{equation}
Finally, by using Equation (\ref{eq:div}) to eliminate the divergence $\nabla\cdot\delta {\textbf{\textit{v}}}$ from the right-hand side of Equation (\ref{eq:p_ast2}) to obtain the ordinary differential equation for $\delta p_\mathrm{tot}$.

\end{document}